\documentclass[aps,twocolumn,showpacs,preprintnumbers,amsmath,amssymb]{revtex4}

\usepackage{graphicx}
\usepackage{dcolumn}
\usepackage{bm}

\newcommand{\gfrac}[2]{\genfrac{}{}{}{}{#1}{#2}}
\newcommand{\thp}{$\Theta^+$}

\begin{document}


\title{JLab: Probing Hadronic Physics with Electrons and Photons}

\author{Elton S. Smith}
\affiliation{Thomas Jefferson National Accelerator Facility, Newport News, Virginia 23606, USA}

\date{\today}

\begin{abstract}
Precision measurements of the structure of nucleons and nuclei
in the regime of strong interaction QCD are now possible
with the availability of high current polarized electron beams, 
polarized targets, and recoil polarimeters, in conjunction 
with modern spectrometers and detector instrumentation.
The physics at JLab will be highlighted using two recent 
measurements of general interest. The ratio of the proton
electric to magnetic form factors indicates the importance
of the role of angular momentum in the structure of the nucleon.
The existence of 5-quark configurations in the ground state
wavefunctions of hadrons is confirmed by a narrow peak
attributed to an exotic baryon with strangeness S=+1.
These and other examples will be used to illustrate the capabilities
and focus of the experimental program at JLab.
\end{abstract}

\maketitle

\section{Introduction to JLab}

The Continuous Electron Beam Accelerator Facility (CEBAF) at the 
Thomas Jefferson National Accelerator Facility (Jefferson Lab) is devoted to the 
investigation of the electromagnetic structure of mesons, nucleons, 
and nuclei using high energy and high duty-cycle 
electron and photon beams. 

CEBAF is a superconducting electron accelerator 
with an initial maximum energy of 4~GeV and 100~\% duty-cycle.  Three 
electron beams with a maximum total current of 200~$\mu A$ can be used 
simultaneously for electron scattering experiments in the experimental 
areas, Halls A, B, and C. The accelerator design concept 
is based on two parallel 
superconducting continuous-wave linear accelerators joined by magnetic 
recirculation arcs. The accelerating structures are five-cell 
superconducting niobium cavities with a nominal average energy gain of
5~MeV/m. The accelerator performance has met all design goals, achieving
5.7 GeV for physics running, and delivering high quality beams with intensity
ratios exceeding 10$^6$:1. The electron beam is
produced using a strained GaAs photocathode delivering polarized electrons
(P$_{e}~\geq$ 75\%) simultaneously to all three halls.

Three experimental areas are available for simultaneous
experiments, the only restriction being that the beam energies have to be 
multiples of the single pass energy. The halls contain
complementary equipment which cover a wide range of physics topics: Hall A
has two high resolution magnetic spectrometers with 10$^{-4}$ 
momentum resolution in a 10\% momentum bite, and a solid angle of 
8~msr. Hall B houses the large 
acceptance spectrometer, CLAS \cite{clas}. Hall C uses a combination of a high momentum
spectrometer ($10^{-3}$ momentum resolution, 7~msr solid angle and maximum
momentum of 7 GeV/c) and a short orbit spectrometer. 

To illustrate the
physics which is being addressed at Jefferson Lab, we have chosen
two topics of current interest: measurements of the electric form
factor of the proton in Hall A, and the observation of an exotic S=+1
baryon with the CLAS.

\section{The shape of the proton at high $Q^2$}

Electron scattering is the tool of choice for the precise investigation of the
spatial structure of nucleons and nuclei. The precision arises from the well-known
characteristics of the electromagnetic interaction. By varying the momentum 
transferred from the electron to the target
for fixed excitation energy, we can directly map out the charge and current
densities, and the transition densities associated with its excitation
\cite{brash}. In the non-relativistic limit and for small four-momentum
transfer squared, $Q^2$, the electric ($G_{Ep}$) and magnetic ($G_{Mp}$) form 
factors are given by the Fourier transforms of the charge and
current distributions in the nucleon. As  $Q^2$ increases, the proton exhibits
its internal structure as a multi-body relativistic system of quarks and gluons.

\begin{figure}[t]
\begin{center}
\includegraphics[width=8cm,bb=91 52  548  711,clip=true]{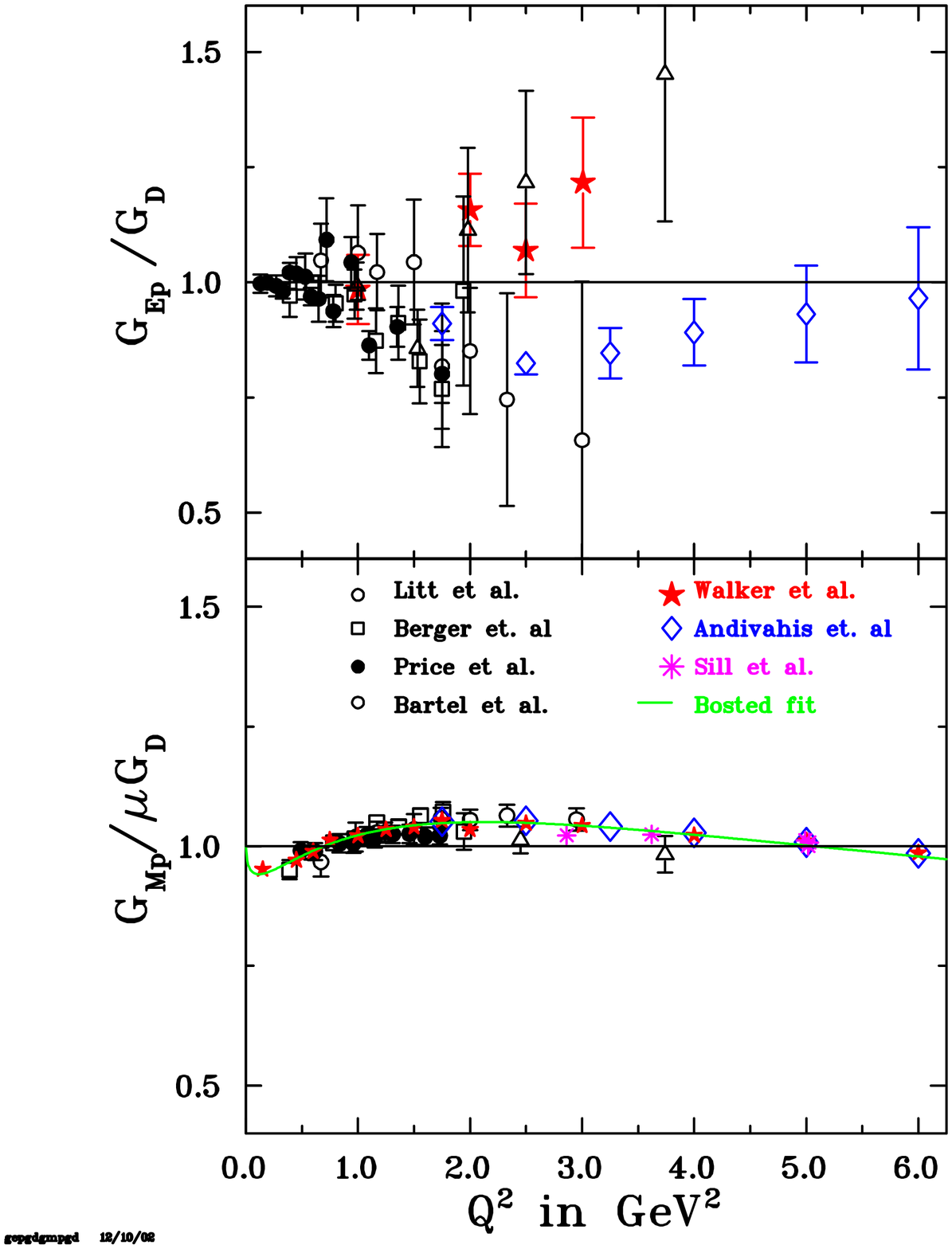}
\caption{\label{fig:fig1} World data prior to 1998 for $G_{Ep}/G_{D}$ (top)
and $G_{Mp}/\mu_p G_{D}$ (bottom).}
\end{center}
\end{figure}

The unpolarized elastic $ep$ cross section is given by:
\begin{eqnarray}
\gfrac{d\sigma}{d\Omega} & = & \gfrac{\alpha^2\cos^2{\frac{\theta_e}{2}}} 
{4 E_e^2 \sin^4{\frac{\theta_e}{2}}} \gfrac{E_{e'}}{E_e}  
\genfrac{(}{)}{}{}{1}{1+\tau}
\left[G^2_{Ep} + \frac{\tau}{\epsilon} G^2_{Mp} \right], \label{eq:eq1}
\end{eqnarray}  
where $E_e$ is the beam energy, $E_{e'}$ and $\theta_e$ are the energy and angle of the
scattered electron, the polarization of the virtual photon is 
$\epsilon = [1 + 2(1+\tau) \tan^2\frac{\theta_e}{2} ]^{-1}$,
and $\tau = Q^2/4M^2_p$ is the four momentum scaled to the proton mass.
The Rosenbluth method uses Eq.\,(\ref{eq:eq1}) to determine the individual
contributions from $G_{Ep}$ and $G_{Mp}$ by using their kinematic dependence 
on $\epsilon$ at fixed $Q^2$. The world data for the electric and magnetic
form factors prior to 1998 is shown in Fig.\,\ref{fig:fig1}.
The form factors are found empirically to approximately
follow the dipole form $G_{D}$ which is used as a reference:
\begin{eqnarray}
G_{Ep} \sim \gfrac{G_{Mp}}{\mu_p} & \sim & G_{D} = \gfrac{1}{\left( 1+\frac{Q^2}{0.71}\right)}
\end{eqnarray}  
In the non-relativistic limit, the dipole shape corresponds to
an exponential change distribution.

At high $Q^2$ the cross section is dominated
by the magnetic term $G_{Mp}$, which makes the determination of the 
electric form factor $G_{Ep}$ by the Rosembluth method increasingly difficult. 
 
Measurements of the ratio of $G_{Ep}/G_{Mp}$ have been completed
at Jefferson Lab for $Q^2$ between 0.5 and 5.6 GeV$^2$ 
by measuring the polarization of the recoil proton in
$\vec{e} p \rightarrow e \vec{p}$ scattering \cite{jones,gayou}.
The scattering of longitudinally polarized electrons on unpolarized protons
results in a transfer of polarization to the recoil proton with two
components in the scattering plane: $P_t$ is perpendicular and $P_l$ is 
the parallel to the proton momentum. The polarization of the proton is
determined using a polarimeter with a graphite or CH$_2$ analyzer
located at the focal plane of the proton spectrometer.
The ratio of electric to magnetic form 
factors is directly proportional to the ratio of polarizations:
\begin{eqnarray}  
\gfrac{G^p_E}{G^p_M} & = & -\gfrac{P_t}{P_l} \gfrac{E_e + E_{e'}}{2 M_p} 
\tan \left(\frac{\theta_e}{2}\right)
\end{eqnarray} 
In first order the polarization normal to the
scattering plane is zero and can serve as a systematic check.

Since the ratio $G^p_E/G^p_M$ is accessed directly, these experiments
are able to carefully control their their systematic uncertainties.
For example, the ratio is independent of the electron beam polarization
and the analyzing power of the polarimeter. Also, detailed knowledge of the
spectrometer acceptances are not needed; the dominant systematic error comes
from uncertainties in calculating the transport of the spin through the magnet.

\begin{figure}[t]
\begin{center}
\includegraphics[width=8cm,clip=true]{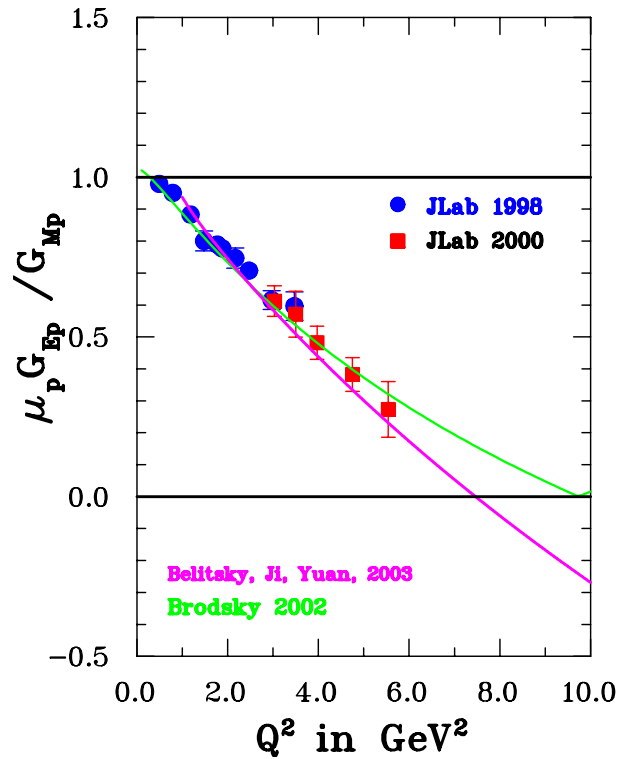}
\caption{\label{fig:fig2} JLab measurements for the ratio
$\mu_p G_{Ep}/G_{Mp}$ as a function of $Q^2$, determined by
measuring the recoil proton polarization. The curves are described in
the text.}
\end{center}
\end{figure}

The results of these measurements are shown in Fig.\,\ref{fig:fig2}
as the ratio $\mu_p G_{Ep}/G_{Mp}$. The plot demonstrates the somewhat
surprising result that the form factor ratio decreases linearly with increasing $Q^2$,
and will cross zero at $Q^2\approx7.5 GeV^2$ if the trend continues.
The data have motivated a flurry of theoretical activity. The leading-order
pQCD predicts that the ratio of Pauli to Dirac form factors should scale as
$F_2/F_1 \sim 1/Q^2$. If logarithmic
corrections are included to the leading-order prediction, 
the ratio behaves more like $F_2/F_1 \sim 1/Q$ \cite{brodsky,belitsky}
and the calculations reproduce the data by adjusting appropriate parameters
as shown in Fig.\,\ref{fig:fig2}. The behavior of the data 
requires that the quarks inside the proton carry
orbital angular momentum \cite{ralston,miller}.   

These new data on  $G_{Ep}$ have motivated discussions regarding their
physical interpretation in terms of the shape of the proton \cite{miller2,belitsky2,ji}.
The calculations have been used to produce images of the proton
by selecting specific configurations of the quark momenta and spins.
When viewed through these ``color'' filters to select specific quark
configurations, the shape of the proton becomes non-spherical.
We await new data to probe deeper into the structure of the proton. The
experiment is approved to extend measurements up to a momentum transfer 
of 9 GeV/c$^2$ \cite{perdrisat}.

\section{Pentaquarks}
  
The question of which color singlet configurations exist in nature lies at the 
heart of strong interaction QCD. Until recently all experimental
evidence indicated that mesons were ($q\bar{q}$) bound states and
the valence structure of baryons was ($qqq$). In the baryon sector,
it is natural to ask whether a 5-quark 
configurations exists where the $\bar{q}$ has a different flavor than 
(and hence cannot annihilate with) the other four quarks. A baryon with the 
exotic strangeness quantum number $S=+1$ is a natural candidate 
for a pentaquark state, because such a state has a minimal
5-quark ($qqqq\bar{q}$) configuration.
Such states are not forbidden \cite{strottman,lipkin}, 
and definite evidence of pentaquark states would be an important 
addition to our understanding of QCD. 

Pentaquark states have been studied both theoretically and 
experimentally for many years \cite{pdg86}. Most recently, 
symmetries within the chiral soliton model were used by 
Diakonov, Petrov and Polyakov \cite{diakonov} 
to predict an anti-decuplet 
of 5-quark resonances with spin and parity $J^\pi = \frac{1}{2}^+$. 
The lowest mass member, now called the \thp, is an isosinglet with valence quark 
configuration $uudd\bar{s}$ giving strangeness $S=+1$ 
with a predicted mass of approximately 1.53 GeV/c$^2$ and a 
width of $\sim 0.015$ GeV/c$^2$. These definite predictions
have prompted experimental searches to focus attention in this
mass region.

This paper describes the evidence for a narrow $S=+1$ baryon
observed in the reaction $\gamma d \rightarrow K^+ K^- p (n)$ using the 
CLAS detector in Hall B \cite{clastheta}. However, 
several other experiments have reported evidence for a state at the
same mass. 
The first observation of an $S=+1$ baryon was reported by the 
LEPS collaboration at the SPring-8 facility in Japan
at a mass of 1.54 GeV/c$^2$, decaying to $n K^+$ with a FWHM less
than $0.025$ GeV/c$^2$ \cite{nakano}.
Confirming evidence has also come from the DIANA collaboration at ITEP \cite{dolgolenko} 
in the $K^0 p$ mass spectrum, the SAPHIR collaboration 
in photoproduction on a proton \cite{SAPHIR}, and most recently
a peak in the $p K^0_s$ system was reported in neutrino and antineutrino
interactions on nuclei \cite{neutrino}.
 
\begin{figure}[t]
\includegraphics[width=1.\columnwidth]{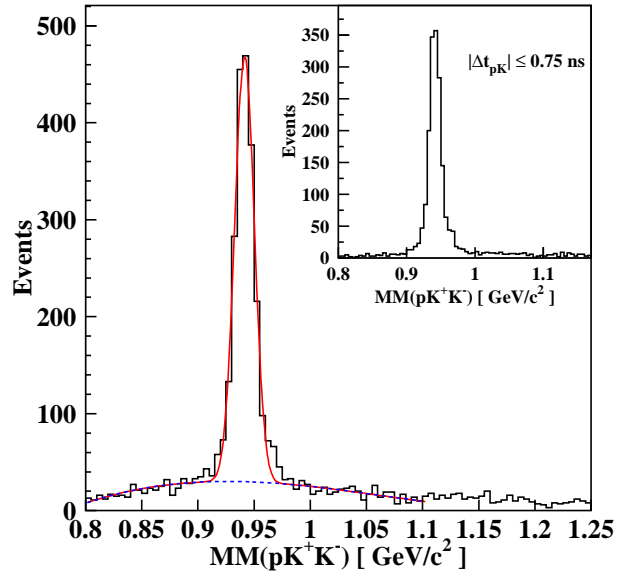}
\caption{ Missing mass spectrum for the $\gamma d \rightarrow pK^+K^- X$ 
reaction, after timing cuts to identify the charged particles and 
the coincident photon, which shows a peak at the neutron mass. 
There is a small, broad background from misidentified particles 
and other sources.  The inset shows the neutron peak with a 
tighter requirement on the timing between the proton and kaons.
}
\label{fig:fig3}
\end{figure}

The CLAS data was taken with a photon beam which was produced by
2.474 and 3.115 GeV electrons incident on a bremsstrahlung 
radiator of thickness $10^{-4}$ radiation lengths, giving a 
tagged photon flux of approximately $4\times 10^6$ $\gamma$'s per
second. The photons were incident on a 10-cm long
liquid-deuterium target.
The event trigger required a single charged track in CLAS 
in coincidence with a hit in the tagging spectrometer.
The momentum of charged particles were reconstructed using magnetic
analysis and their mass determined using time-of-flight techniques.
The analysis selected events with a detected proton, 
$K^+$ and  $K^-$ in the final state, all of which originated
from the same beam bucket.
The detection of all three charged particles allows complete
determination of the reaction and therefore Fermi
motion in the target plays no role in the analysis.
The missing mass ($MM$) of the selected events is plotted 
in Fig.~\ref{fig:fig3}, which shows clear peak at the 
the neutron mass. A fit to the distribution (solid line) yields 
a mass resolution of $\sigma = 0.009$ GeV/c$^2$. 
Events within $\pm 3\ \sigma$ of the neutron peak were 
kept for further analysis.  The background in this region 
is about 15\% of the total, mostly from pions 
that are misidentified as kaons. A cleaner spectrum can
be obtained by applying tighter timing cuts (as shown in the inset), 
but at the expense of reducing the signal.

Several additional cuts were made to optimize the selection of the final sample.
First we required that the reconstructed neutron momentum
be greater than 0.08 GeV/c$^2$. This selection enhances the 
number interactions where the neutron participates
to produce a \thp, and is not a spectator. There are
several known resonances which result in the same final state
and we explicitly removed the two strongest, the $\phi$ meson at 1.02 GeV/c$^2$
and the $\Lambda$(1520). A final cut removed events with
$K^+$ whose momenta exceeds 1.0 GeV/c, which are associated
with invariant masses of the $nK^+$ system above $\sim 1.7$ GeV/c$^2$.

\begin{figure}[t]
\includegraphics[width=1.\columnwidth]{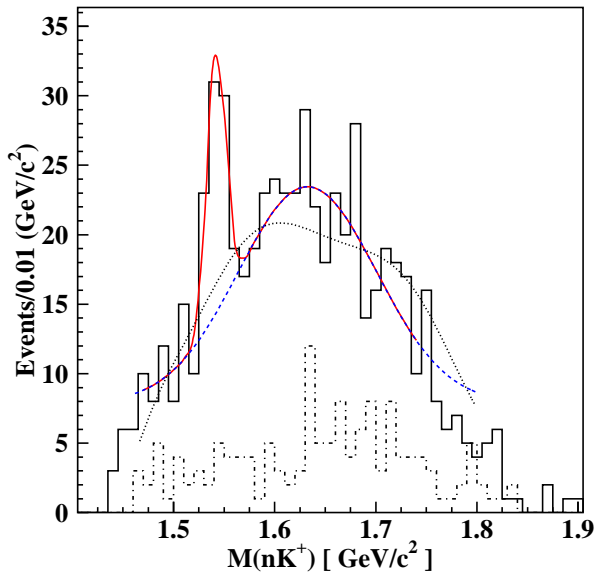}
\caption{ Invariant mass of the $nK^+$ system, which has strangeness 
$S=+1$, showing a sharp peak at the mass of 1.542 GeV/c$^2$.
The dashed-dotted histogram shows the spectrum of events 
associated with $\Lambda$(1520) production. See text for explanation
of the background shapes. 
}
\label{fig:fig4}
\end{figure}

The final $nK^+$ invariant mass spectrum, $M(nK^+)$, 
is shown in Fig.\,\ref{fig:fig4}. The spectrum of events removed by the
$\Lambda$(1520) cut is also shown in Fig.\,\ref{fig:fig4} by the dashed-dotted
histogram, and does not appear to 
be associated with the peak at 1.54 GeV/c$^2$. 
The number of events in the
peak was estimated using several assumptions for the shape
of the background. The solid line fit in the figure uses
an empirical Gaussian plus constant term for the background
(dashed line). This fit determines 43 counts in the peak at a 
mass of $1.542 \pm 0.005$ GeV/c$^2$ with a width (FWHM) of 
0.021 GeV/c$^2$. The dotted line shows a background shape based on
a linear combination of 4-body phase space and 3-body phase space 
of the $pK^+K^-$ final state ($K^+K^-$ in s-wave). The phase space
distributions were generated using a GEANT-based Monte Carlo 
followed by the same reconstruction package as the real data.  

To determine the sensitivity of our experiment, which depends
on the actual shape of the background, ten combinations
of cut placements and fitting functions were tried. 
The estimated statistical significance in those ten cases ranges from
$4.6 \sigma$ to $5.8 \sigma$, which we use to derive the conservative
estimate for the statistical significance of our result of $5.2 \pm
0.6\ \sigma$. 

In summary, evidence is mounting for the existence of a baryon with
a minimum content of 5-quarks. However, the properties of 
this $S=+1$ state, such as spin, isospin and parity,  still need to
be determined before it can be conclusively identified 
with the \thp predicted in Ref.~\cite{diakonov}. In addition to this 
baryon, we expect a whole families of pentaquarks which
promise to rewrite our understanding of baryon spectroscopy. The implications
of this discovery are being evaluated continuously in the literature \cite{maltman}
and we can only expect more surprises in the near future.

\begin{acknowledgments}
I would like to thank R.V. Ribas, and all organizers of the
conference, for their gracious hospitality. I would also like
to thank C.F. Perdrisat, V. Punjabi and E. Brash for their
assistance in preparation of the material on the proton electric
form factor, and to S. Stepanyan for help in the preparation of
materials on the pentaquark. The
Southeastern Universities Research Association 
(SURA) operates the Thomas Jefferson National Accelerator Facility 
for the U.S. Department of Energy under contract DE-AC05-84ER40150.
\end{acknowledgments}


\begin{thebibliography}{99}
\bibitem{clas} B.A. Mecking {\it et al.}, Nucl. Instr. Meth. A {\bf 503}, 
        513 (2003).
\bibitem{brash} E.J. Brash, ``Nucleon Electromagnetic Form Factors,'' p. 256
Baryons 2002, Jefferson Lab, March 3-8, 2002.
\bibitem{jones} M.K. Jones {\it et al.}, Phys. Rev. Lett. {\bf 84}  1398 (2000).
\bibitem{gayou} O. Gayou {\it et al.}, Phys. Rev. Lett. {\bf 88}  092301 (2002).
\bibitem{brodsky} S.J. Brodsky, ``Perspectives on Exclusive Processes in QCD,''
Workshop on Exclusive Processes at High Momentum Transfer, Jefferson Lab, May 15-18, 2002
hep-ph/0208158.
\bibitem{belitsky} A.V. Belitsky, X. Ji and F. Yuan, Phys. Rev. Lett. {\bf 91} 092003 (2003)
 hep-ph/0212351.
\bibitem{ralston} P. Jain and J.P. Ralston, hep-ph/0306194 (2003).
\bibitem{miller} G.A. Miller and M.R. Frank, Phys. Rev. C {\bf 65}, 065205 (2002). 
\bibitem{miller2} G.A. Miller, ``The Shape of the Proton,'' nucl-th/0304076 v2 (2003).
\bibitem{belitsky2} A.V. Belitsky, X. Ji and F. Yuan, ``Quark Imaging in the Proton
Via Quantum Phase-Space Distributions,'' hep-ph/0307383 (2003).
\bibitem{ji} X. Ji, Phys. Rev. Lett. {\bf 91}, 062001-1 (2003).
\bibitem{perdrisat} For a recent review of these measurements, see
V. Punjabi and C.F. Perdrisat, nucl-ex/0307001 (2003), and references therein.

\bibitem{strottman} D. Strottman, Phys. Rev. D {\bf 20}, 748 (1979).
\bibitem{lipkin} H.J. Lipkin, Nucl. Phys. {\bf A625}, 207 (1997).
\bibitem{pdg86} Particle Data Group, Phys. Lett. {\bf 170B}, 289 (1986).
\bibitem{diakonov} D. Diakonov, V. Petrov, and M. Polyakov, 
        Z. Phys. A {\bf 359}, 305 (1997).
\bibitem{clastheta} S. Stepanyan (The CLAS Collaboration), arXiv:hep-ex/0307018
\bibitem{nakano} T. Nakano (The LEPS Collaboration) {\it et al.}, Phys. Rev. Lett. {\bf 91},
012002 (2003) [arXiv:hep-ex/0301020] 
\bibitem{dolgolenko} V.V. Barmin  {\it et al.} (The DIANA Collaboration),
Phys. At. Nucl, {\bf 66}, 1715 (2003) [arXiv:hep-ex/0304040]
\bibitem{SAPHIR} J. Barth {\it et al.} (The SAPHIR Collaboration), arXiv:hep-ex/0307083
\bibitem{neutrino} A.E. Asratyan {\it et al.}, submitted to Yad. Fiz.,
arXiv:hep-ex/0309042.
\bibitem{maltman} For a review of the phenomenology and models, see
B. Jennings and K. Maltman, hep-ph/0308286

\end{thebibliography}
\end{document}